\documentclass{elsart}
\usepackage{epsfig}

\begin{document}
\begin{frontmatter}
\title{An x-ray detector using PIN photodiodes for the axion helioscope}
\author[ICRR]{T. Namba}
\author[ICEPP]{Y. Inoue}
\author[ICRR]{S. Moriyama}
\author[UT]{M. Minowa\corauthref{me}}
\address[UT]{Department of Physics and Research Center for the Early Universe(RESCEU), School of Science, University of Tokyo,
7-3-1, Hongo, Bunkyo-ku, Tokyo 113-0033}
\address[ICEPP]{International Center for Elementary Particle Physics, 
University of Tokyo, 7-3-1, Hongo, Bunkyo-ku,
Tokyo 113-0033, and RESCEU}
\address[ICRR]{Institute for Cosmic Ray Research, University of Tokyo, Kashiwa,
Chiba 277-8582, Japan}
\corauth[me]{Corresponding author\ead{minowa@phys.s.u-tokyo.ac.jp}}

\begin{abstract}
An x-ray detector for a solar axion search was developed.  The
detector is operated at 60K in a cryostat of a superconducting
magnet.  Special care was paid to microphonic noise immunity and
mechanical structure against thermal contraction.  The detector
consists of an array of PIN photodiodes and tailor made preamplifiers.  
The size of each PIN photodiode is $11\times
11\times 0.5\ {\rm mm^3}$ and 16 pieces are used for the detector.
The detector consists of two parts, the front-end part being operated 
at a temperature of 60K and the main
part in room temperature.
Under these circumstances, the detector achieved $1.0$ keV resolution in FWHM,
$2.5$ keV threshold
and $6\times 10^{-5}\ {\mathrm counts\ sec^{-1}\ keV^{-1}\ cm^{-2}}$ background
level.
\end{abstract}

\end{frontmatter}

\section{Axion helioscope}

The axion is a light neutral pseudoscalar particle yet to be discovered.
It was introduced to
solve the strong {\it CP} problem\cite{axion}.
The axion would
be produced in the solar core through the Primakoff effect 
if it has enough coupling to photons\cite{solar_axion}. 
It has an energy spectrum very similar to that
of black body radiation photons with an average energy of about 4 keV.
It can be converted back to an x-ray in a strong magnetic field in the
laboratory by the inverse process.  We search for
such x-rays coming from the direction of the sun with 
a newly developed instrument called axion helioscope.

\begin{figure}[htb]
\begin{center}
\epsfysize=15cm
\epsfbox{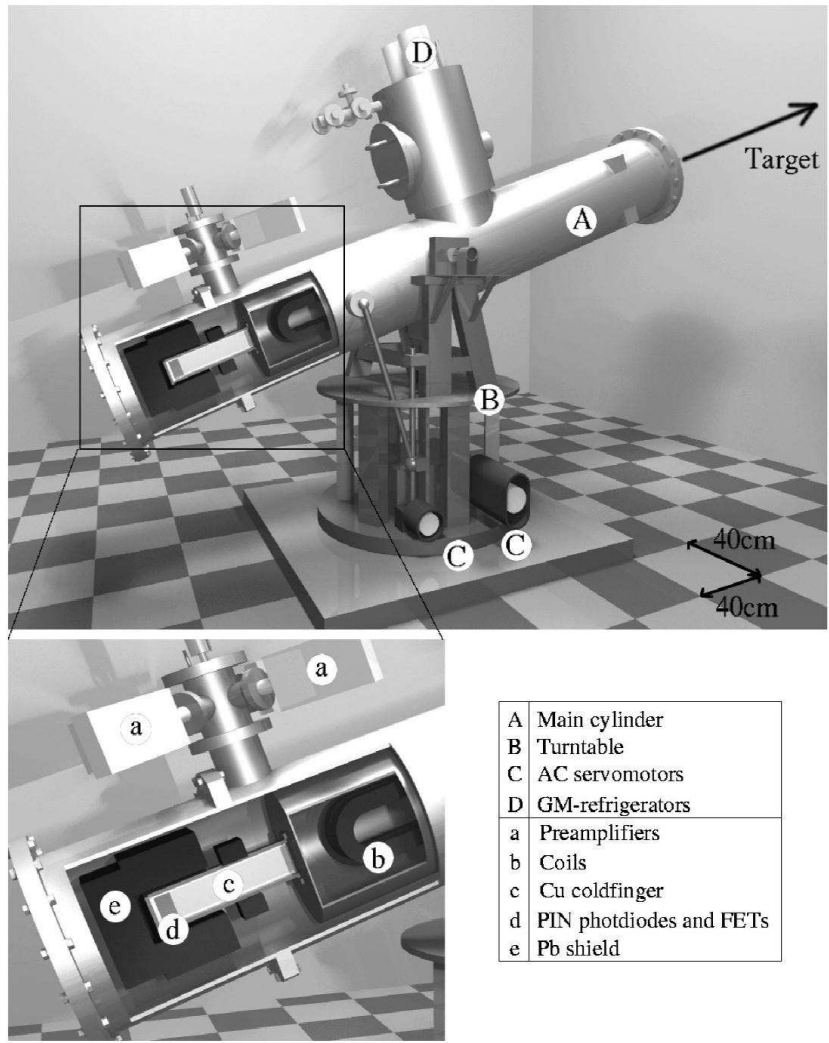}
\caption{Schematic view of the axion helioscope and the tail part of the cylinder.}
\label{sumico_fig}
\end{center}
\end{figure}

A schematic view of the helioscope is shown in Fig. \ref{sumico_fig}.
It consists of three parts, a tracking system,
superconducting coils and x-ray detectors.
The tracking system supports and drives a 3-m long cylinder of 
the helioscope to track the sun.
The superconducting coils and the x-ray detector are mounted in the cylinder.
The coils are cooled by Gifford McMahon refrigerators directly by conduction.
The aperture volume between the coils is $2300 {\rm (L)}\times 92{\rm (H)}\times 
20{\rm (W)}\ {\rm mm^3}$ and
the magnetic field strength in the aperture is 4\ T.
The details of this cryogen-free superconducting magnet 
is described elsewhere\cite{magnet}.

By using the helioscope, we performed a series of axion search 
experiments\cite{results} 
and obtained a limit to the coupling constant of the axion to photons,
$g_{{\rm a}\gamma \gamma} < 6.0$ -- $9.6\times 10^{-10}\ {\rm GeV^{-1}}$
for $m_{\rm a}< 0.26\ {\rm eV}$.

We report in this paper 
on the development of the x-ray detector made of PIN photodiode 
operated in a low temperature environment.

\section{PIN photodiode}\label{detector_section}

The x-ray detector for the axion helioscope is required to have following
features.

\begin{itemize}
\item large area

Rare event search needs large sensitive area.

\item tolerance to magnetic field

The detector is operated near the superconducting coils,
and operated under 10$^{-2}$\, T magnetic field.

\item low energy threshold and moderately good resolution

Since the energy spectrum extends to zero energy, the threshold needs be low enough.
The higher resolution also helps but needs not be excellent because the spectrum has 
no sharp structures.

\item low-temperature operation

No thermal insulation is possible between the detector and the superconducting
magnet while keeping good x-ray transmission.

\item high efficiency for low energy x-rays

A thick window would stop low energy x-rays and lose low energy
portion of the spectrum.

\item low background

In a rare event search,
the remaining background rate limits the sensitivity.



\end{itemize}

We found silicon PIN photodiodes with thick depletion 
layers satisfy the above requirements\cite{prev_paper}.
The PIN photodiode we use is Hamamatsu S3590-06-SPL, $11\times 11\ {\rm mm^2}$
in size and $500\ {\rm \mu m}$ in thickness.
It is originally a windowless silicon PIN photodiode but
it has also high efficiency to x-rays of energy less than 10 keV.
To cover the full aperture of the superconducting magnet,
an array of 16 PIN photodiodes was used. 

\begin{figure}[htb]
\begin{center}
\epsfysize=6cm
\epsfbox{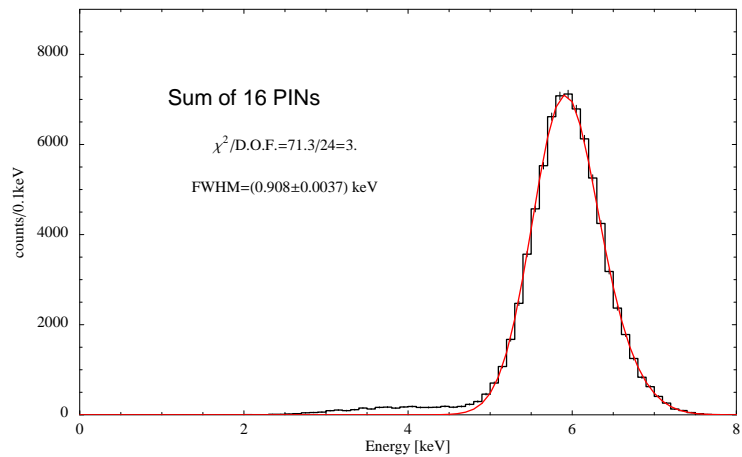}
\caption{Spectrum of 5.9\,keV Mn-x rays measured by 16 PIN photodiodes.}
\label{calib_spec_fig}
\end{center}
\end{figure}

As measured with the calibration source, the energy resolution of
the PIN-photodiode detector is 0.9 to 1.0 keV in FWHM at 5.9 keV.  
The measured spectrum is shown in Fig. \ref{calib_spec_fig}
The energy threshold is 2.5 keV.

\begin{figure}[htb]
\begin{center}
\epsfysize=10cm
\epsfbox{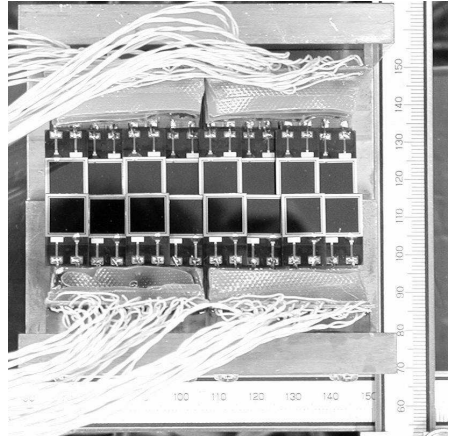}
\caption{The array of the PIN photodiodes as seen from the
incident direction. The figures marked on the scale are in millimeters.}
\label{array_fig}
\end{center}
\end{figure}

Although the PIN photodiode is of windowless type,
it has an aluminum electrode 
along four sides of the surface and they block the x-rays.
To get rid of this inefficient section, the PIN photodiodes were arranged
to overlap on their electrodes each other
(Fig. \ref{array_fig}).

Besides the electrode, 
there exists a slight thickness of inefficient silicon dioxide in front of 
the depletion layer of the PIN photodiode.
The thickness of the insensitive layer was estimated by measuring the inefficiency 
of
$5.9$ keV x-rays
from ${}^{55}{\rm Fe}$ source, and was found to be
less than $6.1\ {\rm \mu m}$ at the 95 \% confidence level.
This means that the efficiency is more than 0.2 for 3 keV x-ray and 0.92 for 10keV.

A surface-scan measurement guaranteed that the efficiency
is uniform in $9\times 9\ {\rm mm^2}$ region inside the electrode.

\section{Assembly}

The x-ray detector is divided into two parts, the cold head part and the main
part in room temperature.
The head part of the x-ray detector is mounted on a 60 K copper plate and
surrounded by shields to reduce the radioactive
background from the environment(Fig. \ref{sumico_fig}) .
In the head part, PIN photodiodes and front-end circuits are assembled.

\begin{figure}[htb]
\begin{center}
\epsfysize=7cm
\epsfbox{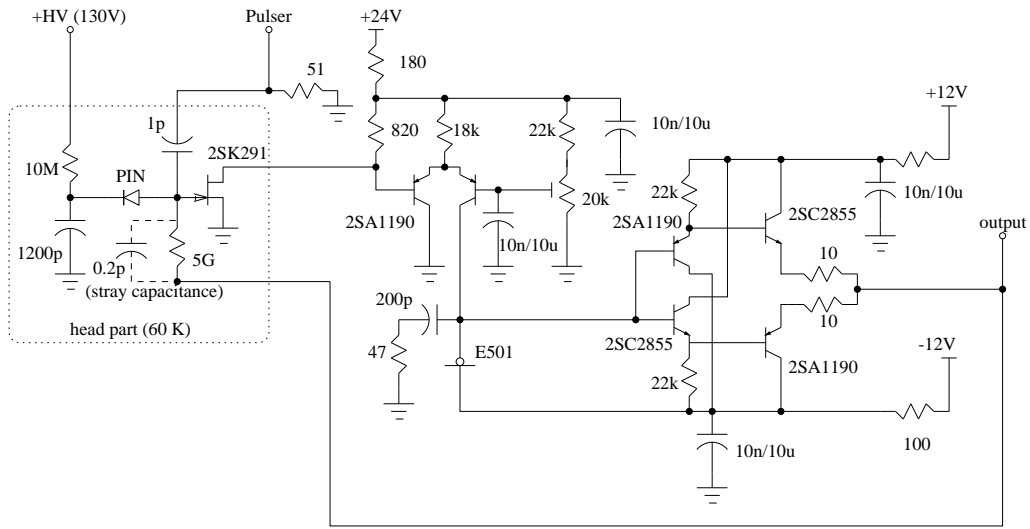}
\caption{Schematic diagram of the amplifier.}
\label{schematic_fig}
\end{center}
\end{figure}

Just beside the array of the PIN photodiodes,
there are front-end circuits of charge-sensitive preamplifiers.
The schematic diagram of the amplifier is shown in Fig. \ref{schematic_fig}.
In the front-end circuits are
junction field-effect transistors (JFETs, HITACHI 2SK291),
feedback resisters (5 G$\Omega$), 130 V inverse bias voltage buffers and
pulser inputs.
The residual parts of the preamplifiers are
mounted outside of the cylinder at the room temperature.
Also in the room temperature part are high-pass filters 
and booster amplifiers, which we will mention later. 

The head part of the x-ray detector is mounted on 60 K plate in order not to
radiate much thermal energy onto the 4 K superconducting coils.
The low temperature is also useful 
to reduce the thermal noise in the front-end circuits,
because the electric noises of the PIN photodiodes and circuits decrease
as the temperature drops.
However, the JFET does not function at such a low temperature.
Therefore, the JFET was warmed up to a temperature around 130 K 
by a small electric heater attached on it.
Between the front-end circuit and the 60 K plate, 
6 mm thick acrylic plastic sheets
are inserted for the thermal insulation, whose impedance saves necessary heat
to warm the JFET up to the working temperature.

For the calibration of assembled PIN photodiodes, an x-ray source of 
$^{55}$Fe is installed in front of them.  
The source can be retracted behind the shield during 
the run so that the storage of the check source inside the cryostat does not 
increase the background level. 

\section{Background and noise}
Since the PIN photodiode is made of silicon whose atomic
number is small, it has relatively low photoelectric efficiency
for high energy photons.
Therefore the detector should be properly shielded lest
the Compton scattering of photons form continuous
backgroud at the low energy region.

The radioactive background consists of two parts, external and internal.
The external component is reduced by the shield
shown in Fig. \ref{sumico_fig}.
The PIN photodiodes are surrounded by 10 cm
thick lead, (partly mounted on room temperature part and partly on 60 K),
1 cm oxygen free copper and 0.3--1 cm acrylic plastic sheets.  
The copper is also used as 60 K thermal radiation shield and as a cold finger to
cool the x-ray detectors.

On the other hand, one of the possible internal 
radioactive source is the ceramic
commonly used for the base of the PIN photodiode.
The ceramic has usually ${}^{238}{\rm U}$ and ${}^{232}{\rm Th}$ chain
contaminations which amount to 
about $10^{-2}$--$10^{-3}\ {\rm Bq}$ per PIN photodiode.
To get rid of these radioactivities, the base material was changed.
The new bases are made of flexible polyimide sheets. 
Their size is $11\times 16\ {\rm mm^3}$ which just fits to
our PIN photodiode array.
With this new bases the background rate is reduced to
$6\times 10^{-5} {\rm counts\ sec^{-1}\ keV^{-1}\ cm^{-2}}$.

\begin{figure}[htb]
\begin{center}
\epsfysize=3cm
\epsfbox{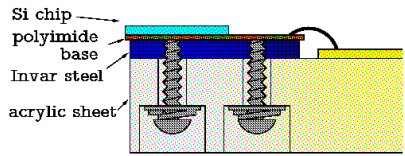}
\caption{PIN photodiode mounted on an Invar steel sheet.}
\label{head_fig}
\end{center}
\end{figure}

The superconducting coils of the helioscope is
directly cooled by Gifford-McMahon refrigerators,
and its continuous back-and-forth movement
causes microphonic noises of the detector.
To reduce the microphonic noises,
connections between the PIN photodiodes and the front-end circuits
are kept as short and rigid as possible.
As is shown in Fig. \ref{head_fig} the bases of the PIN photodiodes are fixed on
Super Invar steel sheets by epoxy resin.
The Super Invar sheets are then screwed on the acrylic sheets.
Invar steel has as small thermal contraction coefficient as the silicon and
prevents the silicon from cracking due to the thermal contraction 
of the acrylic sheet.
All parts of the front-end circuit were molded with Stycast-1266 epoxy resin which
is mixed with \#250-mesh ${\rm SiO}_2$ powder with a weight fraction of 55\%
to make its overall contraction coefficient
comparable to the parts in the mold.

Signals from the preamplifiers are amplified by the booster
with a gain of 0--1000 and then the waveform is digitized at
a sampling rate of 10 MHz with a length of 1024 words per event.
The high-pass filters are inserted in front of the boosters
to reduce microphonic noises whose major components reside
below several 100 Hz.
The cut-off frequencies are set to be 3 kHz, since signal
below this frequency has less meaning considering the sampling
duration of $102.4\,\mu s$.

Shaping of the signal is done off-line and
still remaining microphonic events can be removed digitally.

\section{Conclusion}

We developed an x-ray detector for the axion helioscope using a
PIN photodiodes array.
It works at a low temperature of 60 K, with low background, low microphonics and high 
sensitivity for low energy x-rays.

With this system we performed a series of axion search 
experiments\cite{results} .
We are now preparing more massive axion search experiment
with our helioscope by filling buffer gas into the magnetic field region.

\section*{Acknowledgement}
The authors thank F.~Shimokoshi who designed the original version 
of the preamplifier circuit. 

The present research is supported by the Grant-in-Aid for COE research by
Japanese Ministry of Education, Culture, Sports, Science and Technology.

\end{document}